\documentclass[journal,letterpaper]{IEEEtran}

\usepackage{graphicx}
\usepackage{amsmath,amssymb}
\usepackage{balance}
\usepackage{url}

\newcommand{\NML}{{\ensuremath{\text{\emph{nml}}}}}

\newcommand{\mix}{{\ensuremath{\text{\emph{mix}}}}}

\newcommand{\qedsymbol}{\rule{1ex}{1ex}}
\newcommand{\qed}{\mbox{}\hfill\qedsymbol}

\newtheorem{remark}{Remark}

\newcommand{\normal}{{\ensuremath{\mathcal{N}}}}

\newcommand{\calW}{{\ensuremath{\mathcal{W}}}}

\newcommand{\eqdef}{{\ensuremath{\;:=\;}}}

\newcommand{\coeff}{{\ensuremath{c}}}
\newcommand{\indic}{{\ensuremath{\mathbb{I}}}}

\newcommand{\simiid}{{\ensuremath{\overset{\mathit{i.i.d.}}{\sim}}}}

\newtheorem{proposition}{Proposition}

\begin{document}

\title{MDL Denoising Revisited}

\author{Teemu~Roos, Petri~Myllym\"aki,
        and~Jorma~Rissanen
	\thanks{Authors are with the Complex Systems Computation
        Group, Helsinki Institute for Information Technology.
	e-mails: {teemu.roos@cs.helsinki.fi}, 
	{petri.myllymaki@cs.helsinki.fi}, 
	{jorma.rissanen@mdl-research.org}.
	This work was supported in part by the Academy of Finland
        under project \textsc{Minos}, the Finnish Technology Agency under
        projects PMMA and KUKOT, and the IST Programme of the European
        Community, under the PASCAL Network of Excellence,
        IST-2002-506778.
        }}

\markboth{Draft, June 28, 2006}{}

\maketitle

\begin{abstract}
  We refine and extend an earlier MDL denoising criterion for
  wavelet-based denoising. We start by showing that the denoising
  problem can be reformulated as a clustering problem, where the goal
  is to obtain separate clusters for informative and non-informative
  wavelet coefficients, respectively.  This suggests two refinements,
  adding a code-length for the model index, and extending the model in
  order to account for subband-dependent coefficient distributions.  A
  third refinement is derivation of soft thresholding inspired by
  predictive universal coding with weighted mixtures. We propose a
  practical method incorporating all three refinements, which is shown
  to achieve good performance and robustness in denoising both
  artificial and natural signals.
\end{abstract}


\IEEEpeerreviewmaketitle

\section{Introduction}

\PARstart{W}{avelets} are widely applied in many areas of signal
processing~\cite{mallat:1998}, where their popularity owes largely to
efficient algorithms on the one hand and advantages of sparse
wavelet representations on the other.  The sparseness property means
that while the distribution of the original signal values may be very
diffuse, the distribution of the corresponding wavelet coefficients is
often highly concentrated, having a small number of very large values
and a large majority of very small values~\cite{mallat:1989}.  It is
easy to appreciate the importance of sparseness in signal compression,
~\cite{devore:1992,villasenor:1995}. The task of removing noise from
signals, or \emph{denoising}, has an intimate link to data
compression, and many denoising methods are explicitly designed to
take advantage of sparseness and compressibility in the wavelet
domain, see e.g.,~\cite{natarajan:1995}--\nocite{hansen:2000}%
\cite{liu:2001}.

Among the various wavelet-based denoising methods those suggested by
Donoho and
Johnstone~\cite{donoho-johnstone:1994,donoho-johnstone:1995} are the
best known.  They follow the frequentist minimax approach, where the
objective is to asymptotically minimize the worst-case $L^2$ risk
simultaneously for signals, for instance, in the entire scale of
H\"older, Sobolev, or Besov classes, characterized by certain
smoothness conditions.  By contrast, Bayesian denoising methods
minimize the \emph{expected} (Bayes) risk, where the expectation is
taken over a given prior distribution supposed to govern the unknown
true signal ~\cite{vidakovic:1998,ruggeri:1999}.  Appropriate prior
models with very good performance in typical benchmark tests,
especially for images, include the class of generalized Gaussian
densities ~\cite{hansen:2000,chang:2000,moulin:1999}, and
scale-mixtures of Gaussians~\cite{wainwright:2000,portilla:2003} (both
of which include the Gaussian and double exponential densities as
special cases).

A third approach to denoising is based on the minimum description
length (MDL) principle~\cite{rissanen:1978}--\nocite{rissanen:1996}%
\nocite{grunwald:2005}\cite{grunwald:2007}. Several different MDL
denoising methods have been
suggested~\cite{hansen:2000,chang:2000,saito:1994}--%
\nocite{antoniadis:1997}\nocite{krim:1999}\nocite{rissanen:2000}%
\cite{kumar:2006}.  We focus on what we consider as the
most pure MDL approach, namely that of
Rissanen~\cite{rissanen:2000}. Our motivation is two-fold: First, as
an immediate result of refining and extending the earlier MDL
denoising method, we obtain a new practical method with greatly
improved performance and robustness. Secondly, the denoising problem
turns out to illustrate theoretical issues related to the MDL
principle, involving the problem of unbounded parametric complexity
and the necessity of encoding the model class. The study of denoising
gives new insight to these issues.

Formally, the denoising problem is the following. Let $y^n =
(y_1,\ldots,y_n)^T$ be a signal represented by a real-valued column
vector of length $n$.  The signal can be, for instance, a
time-series or an image with its pixels read in a row-by-row order.
Let $\calW$ be an $n \times m$ regressor matrix whose columns are
basis vectors.  We model the signal $y^n$ as a linear combination of
the basis vectors, weighted by coefficient vector $\beta^n =
(\beta_1,\ldots,\beta_m)^T$, plus Gaussian i.i.d.\ noise:
\begin{equation}
\label{eq:model}
y^n = \calW\beta^m + \epsilon^n, \quad \epsilon_i \simiid
\normal(0,\sigma_N^2),
\end{equation}
where $\sigma_N^2$ is the noise variance.  Given an observed signal
$y^n$, the ideal is to obtain a coefficient vector $\tilde\beta^m$
such that the signal given by the transform $\tilde{y}^n = \calW
\tilde\beta^m$ contains the informative part of the observed signal,
and the difference $y^n - \tilde{y}^n$ is noise.

For technical convenience, we adopt the common restriction on
$\calW$ that the basis vectors span a \emph{complete orthonormal}
basis. This implies that the number of basis vectors is equal to the
length of the signal, $m=n$, and that all the basis vectors are
orthogonal unit vectors. There are a number of wavelet transforms
that conform to this restriction, for instance, the Haar transform
and the family of Daubechies
transforms~\cite{mallat:1998,daubechies:1992}.  Formally, the matrix
$\calW$ is of size $n \times n$ and orthogonal with its inverse
equal to its transpose. Also the mapping $\beta^n \mapsto \calW
\beta^n$ preserves the Euclidean norm, and we have Parseval's
equality:
\begin{equation}
\label{eq:parseval}
||\beta^n|| = \sqrt{\langle \beta^n,\beta^n \rangle}
= \sqrt{\langle \calW \beta^n, \calW \beta^n\rangle} = ||\calW \beta^n||.
\end{equation}
Geometrically this means that the mapping $\beta^n \mapsto \calW \beta^n$
is a rotation and/or a reflection.  From a statistical point of
view, this implies that any spherically symmetric density, such as
Gaussian, is invariant under this mapping.  All these properties are
shared by the mapping $y^n \mapsto \calW^Ty^n$.  We call $\beta^n
\mapsto \calW \beta^n$ the inverse wavelet transform, and $y^n
\mapsto \calW^Ty^n$ the forward wavelet transform. Note that in
practice the transforms are not implemented as matrix
multiplications but by a fast wavelet transform similar to the fast
Fourier transform (see~\cite{mallat:1998}), and in fact not even the
matrices need be written down.

For complete bases, the conventional maximum likelihood (least
squares) method obviously fails to provide denoising unless the
coefficients are somehow restricted since the solution
$\tilde\beta^n = \calW^T y^n$ gives the reconstruction $\tilde{y}^n
= \calW \calW^T y^n = y^n$ equal to the original signal, including
noise.  The solution proposed by Rissanen~\cite{rissanen:2000} is to
consider each subset of the basis vectors separately and to choose
the subset that allows the shortest description of the data at hand.
The length of the description is determined by the normalized
maximum likelihood (NML) code length.

The NML model involves an integral, which is undefined unless the
range of integration (the support) is restricted. This, in turn,
implies hyper parameters, which have received increasing attention
in various contexts involving, e.g., Gaussian, Poisson and geometric
models~\cite{rissanen:1996,grunwald:2007,foster:2001}--\nocite{foster:2005}%
\nocite{liang:2004}\cite{derooij:2006}. Rissanen used
renormalization to remove them and to obtain a second-level NML
model. Although the range of integration has to be restricted also
in the second-level NML model, the range for ordinary regression
problems does not affect the resulting criterion and can be ignored.
Roos et al.~\cite{roos:2005} give an interpretation of the method
which avoids the renormalization procedure and at the same time
gives a simplified view of the denoising process in terms of two
Gaussian distributions fitted to informative and non-informative
coefficients, respectively. In this paper we carry this
interpretation further and show that viewing the denoising problem
as a clustering problem suggests several refinements and extensions to
the original method.

The rest of this paper is organized as follows. In Sec.~\ref{sec:mdl}
we reformulate the denoising problem as a task of clustering the
wavelet coefficients in two or more sets with different
distributions. In Sec.~\ref{sec:refined} we propose three different
modifications of Rissanen's method, suggested by the clustering
interpretation.  In Sec.~\ref{sec:results} the modifications are shown
to significantly improve the performance of the method in denoising
both artificial and natural signals. The conclusions are summarized in
Sec.~\ref{sec:conclusions}.

\section{Denoising and Clustering}
\label{sec:mdl}

\subsection{Extended Model}

We rederive the basic model~(\ref{eq:model}) in such a way that there
is no need for renormalization.  This is achieved by inclusion of the
coefficient vector $\beta$ in the model as a variable and by selection
of a (prior) density for $\beta$. While the resulting NML model will
be equivalent to Rissanen's renormalized solution, the new formulation
is easier to interpret and directly suggests several refinements and
extensions.

Consider a fixed subset $\gamma \subseteq \{1,\ldots,n\}$ of the
coefficient indices.  We model the coefficients $\beta_i$ for $i \in
\gamma$ as independent outcomes from a Gaussian distribution with
variance $\tau^2$. In the basic hard threshold version all $\beta_i$
for $i \notin \gamma$ are forced to equal zero. Thus the extended
model is given by
\begin{equation}
\label{eq:extendedmodel}
y^n = \calW\beta^n + \epsilon^n, \quad
\begin{cases}\epsilon_i \simiid \normal(0,\sigma_N^2),\\
  \beta_i \simiid \normal(0,\tau^2), &\mbox{if $i\in\gamma$,}\\
  \beta_i = 0, &\mbox{otherwise.}
\end{cases}
\end{equation}
This way of modeling the coefficients is akin to the so called
\emph{spike and slab} model often used in Bayesian variable
selection~\cite{mitchell:1988,george:1997} and applications to
wavelet-based denoising~\cite{chipman:1997,abramovich:1998} (and
references therein). In relation to the sparseness property
mentioned in the introduction, the `spike' consists of coefficients
with $i \notin \gamma$ that are equal to zero, while the `slab'
consists of coefficients with $i \in \gamma$ described by a Gaussian
density with mean zero. This is a simple form of a scale-mixture of
Gaussians with two components. In Sec.~\ref{sec:subband} we will
consider a model with more than two components.

Let $\coeff^n = \beta^n + \calW^T\epsilon^n$, where
$\calW^T\epsilon^n$ gives the representation of the noise in the
wavelet domain. The vector $\coeff^n$ is the wavelet representation
of the signal $y^n$, and we have $$y^n = \calW\beta^n + \calW
\calW^T\epsilon^n = \calW\coeff^n.$$ It is easy to see that the
maximum likelihood parameters are obtained directly from
\begin{equation}
\label{eq:mlbeta}
   \hat{\beta}_i = \begin{cases} \coeff_i, &\mbox{if $i\in \gamma$,}\\
     0, &\mbox{otherwise.}
\end{cases}
\end{equation}
The i.i.d.\ Gaussian distribution for $\epsilon^n$ in
(\ref{eq:extendedmodel}) implies that the distribution of
$\calW^T\epsilon^n$ is also i.i.d. and Gaussian with the same
variance, $\sigma_N^2$.  As a sum of two independent random
variates, each $\coeff_i$ has a distribution given by the
convolution of the densities of the summands, $\beta_i$ and the
$i$th component of $\calW^T\epsilon^n$.  In the case $i \notin
\gamma$ this is simply $\normal(0, \sigma_N^2)$. In the case $i \in
\gamma$ the density of the sum is also Gaussian, with variance given
by the sum of the variances, $\tau^2 + \sigma_N^2$. All told, we
have the following simplified representation of the extended model
where the parameters $\beta^n$ are implicit:
\begin{equation}
\label{eq:simplermodel}
y^n = \calW\coeff^n, \quad
\coeff_i \simiid
\begin{cases}\normal(0,\sigma_I^2), &\mbox{if $i\in\gamma$,}\\
  \normal(0,\sigma_N^2), &\mbox{otherwise,}
\end{cases}
\end{equation}
where $\sigma_I^2 \eqdef \tau^2 + \sigma_N^2$ denotes the variance of
the informative coefficients, and we have the important restriction
$\sigma_I^2 \geq \sigma_N^2$ which we will discuss more below.


\subsection{Denoising Criterion}
\label{sec:clusteringproblem}

The task of choosing a subset $\gamma$ can now be seen as a
clustering problem: each wavelet coefficient belongs either to the
set of the informative coefficients with variance $\sigma_I^2$, or
the set of non-informative coefficients with variance $\sigma_N^2$.
The MDL principle gives a natural clustering criterion by
minimization of the code-length achieved for the observed signal
(see~\cite{kontkanen:2005}). Once the optimal subset is identified,
the denoised signal is obtained by setting the wavelet coefficients
to their maximum likelihood values~\eqref{eq:mlbeta}; i.e.,
retaining the coefficients in $\gamma$ and discarding the rest, and
doing the inverse transformation. It is well known that this amounts
to an orthogonal projection of the signal to the subspace spanned by
the wavelet basis vectors in $\gamma$.

The code length under the model~\eqref{eq:simplermodel} depends on
the values of the two parameters, $\sigma_I^2$ and $\sigma_N^2$. The
standard solution in such a case is to construct a single
representative model for the whole model class\footnote{Here the
usual terminology where the word `model' has double meaning is
somewhat unfortunate. The term refers to both a set of densities
such as the one defined by Eq.~\eqref{eq:simplermodel} (as in the
`Gaussian model', or the `logistic model'), and a single density
such as the NML model, which can of course be thought of as a
singleton set. Whenever there is a chance of confusion, we use the
term `model class' in the first sense.} such that the representative
model is universal (can mimic any of the densities in the
represented model class). The minimax optimal universal model
(see~\cite{rissanen:2001}) is given by the so called normalized
maximum likelihood (NML) model, originally proposed by
Shtarkov~\cite{shtarkov:1987} for data compression.  We now consider
the NML model corresponding to the extended
model~\eqref{eq:simplermodel} with the index set $\gamma$ fixed.

Denote by $k = k(\gamma)$ the number of coefficients for which $i
\in \gamma$. The NML density under the extended
model~\eqref{eq:simplermodel} for a given coefficient subset
$\gamma$ is defined as
$$
   f_\NML(y^n \;;\; \gamma) := \frac{f(y^n \;;\; \hat\sigma^2_I,
     \hat\sigma^2_N)}{C_\gamma},
$$ where $\hat\sigma^2_I = \hat\sigma^2_I(y^n)$ and $\hat\sigma^2_N =
\hat\sigma^2_N(y^n)$ are the maximum likelihood parameters for the
data $y^n$, and $C_\gamma$ is the important normalizing constant.
The constant $C_\gamma$ is also known as the \emph{parametric
complexity} of the model class defined by $\gamma$.

Restricting the data such that the maximum likelihood parameters
satisfy
$$
\sigma^2_{\min} \leq \hat\sigma_N^2, \hat\sigma_I^2 \leq \sigma^2_{\max},
$$ and ignoring the constraint $\sigma_N^2 \leq \sigma_I^2$, the 
code length 
under the extended model~\eqref{eq:simplermodel} is approximated
by\footnote{We express code lengths in \emph{nats} which corresponds
to the use of the natural logarithm. One nat is equal to $(\ln
2)^{-1}$ bits.}
\begin{multline}\label{eq:onelevel}
  \frac{n-k}{2} \ln \frac{S(y^n)-S_\gamma(y^n)}{n-k}
  + \frac{k}{2} \ln \frac{S_\gamma(y^n)}{k}
  + \frac{1}{2} \ln k(n-k),
\end{multline}
plus a constant independent of $\gamma$, with $S(y^n)$ and
$S_\gamma(y^n)$ denoting the sum of the squares of all the wavelet
coefficients and the coefficients for which $i \in \gamma$,
respectively
(see the appendix for a proof
). The code length formula is very accurate even for small $n$ since it
involves only the Stirling approximation of the Gamma function.

\begin{remark}
The set of sequences satisfying the restriction $\sigma^2_{\min}
\leq \hat\sigma_N^2, \hat\sigma_I^2 \leq \sigma^2_{\max}$ depends on
$\gamma$. For instance, consider the case $n=2$. In a model with
$k=1$, the restriction corresponds to a union of four squares,
whereas in a model with either $k=0$ or $k=2$, the relevant area is
an annulus (two-dimensional spherical shell). However, the
restriction can be understood as a definition of the support of the
corresponding NML model, not a rigid restriction on the data, and
hence models with varying $\gamma$ are still comparable as long as
the maximum likelihood parameters for the observed sequence satisfy
the restriction.
\end{remark}



The  code length obtained is identical to that derived by Rissanen
with renormalization~\cite{rissanen:2000} (note the correction to
the third term of~\eqref{eq:onelevel} in~\cite{lectures}).  The
formula has a concise and suggestive form that originally lead to
the interpretation in terms of two Gaussian
densities~\cite{roos:2005}. It is also the form that has been used
in subsequent experimental work with somewhat mixed
conclusions~\cite{roos:2005,ojanen:2004}: While for Gaussian low
variance noise it gives better results than a universal threshold of
Donoho and Johnstone~\cite{donoho-johnstone:1994} (VisuShrink),
over-fitting occurs in noisy cases~\cite{roos:2005} (see also
Sec.~\ref{sec:results} below), which is explained by the fact that
omission of the third term is justified only in regression problems
with few parameters.



\begin{remark} It was proved in~\cite{rissanen:2000} that the
criterion~\eqref{eq:onelevel} is minimized by a subset $\gamma$
which consists of some number $k$ of the largest or smallest wavelet
coefficients in absolute value. It was also felt that in denoising
applications the data are such that the largest coefficients will
minimize the criterion. The above alternative formulation gives a
natural solution to this question: by the inequality $\sigma_I^2
\geq \sigma_N^2$, the set of coefficients with larger variance,
i.e., the one with larger absolute values should be retained, rather
than \emph{vice versa}.
\end{remark}

\begin{remark}
In reality the NML model corresponding to the extended
model~(\ref{eq:simplermodel}) is identical to Rissanen's
renormalized model only if the inequality $\sigma_I^2 \geq
\sigma_N^2$ is ignored in the calculations (see the appendix).
However, the following proposition (proved in the appendix) shows that
the effect of doing so is independent of $k$, and hence irrelevant.
\end{remark}

\begin{proposition}\label{prop:ignore}
The effect of ignoring the constraint $\sigma_N^2 \leq \sigma_I^2$
is exactly one bit.
\end{proposition}

We can safely ignore the constraint and use the model without the
constraint as a starting point for further developments for the sake of
mathematical convenience.

\section{Refined MDL Denoising}
\label{sec:refined}

\subsection{Encoding the Model Class}
\label{sec:encodemodel}

It is customary to ignore encoding of the index of the model class
in MDL model selection; i.e., encoding the number of parameters when
the class is in one-to-one correspondence with the number of
parameters. One simply picks the class that enables the shortest
description of the data without considering the number of bits
needed to encode the class itself. Note that here we do not refer to
encoding the parameter values as in two-part codes, which are done
implicitly in the so-called `one-part codes' such as the NML and
mixture codes. In most cases there are not too many classes and
hence omitting the code length of the model index has no practical
consequence. When the number of model classes is large, however,
this issue does become of importance. In the case of denoising, the
number of different model classes is as large as $2^n$ (with $n$ as
large as $512 \times 512 = 262,144$) and, as we show, encoding of
the class index is crucial.

The encoding method we adopt for the class index is simple. We first
encode $k$, the number of retained coefficients with a uniform code,
which is possible since the maximal number $n$ is fixed. This part
of the code can be ignored since it only adds a constant to all code
lengths. Secondly, for each $k$ there are a number of different
model classes depending on which $k$ coefficients are retained. Note
that while the retained coefficients are always the \emph{largest}
$k$ coefficients, this information is not available to the decoder
at this point and the index set to be retained has to be encoded.
There are ${n \choose k}$ sets of size $k$, and we use a uniform
code yielding a code length $\ln {n \choose k}$ nats, corresponding
to a prior probability
\begin{equation}
\label{eq:prior}
   \pi(\gamma) =  {n \choose k}^{-1} = \frac{k! (n-k)!}{n!}.
\end{equation}

Applying Stirling's approximation to the factorials and ignoring all
constants wrt.~$\gamma$ gives the final code length formula
\begin{equation}
\label{eq:withchoose}
  \frac{n-k}{2} \ln \frac{S(y^n)-S_\gamma(y^n)}{(n-k)^3}
  + \frac{k}{2} \ln \frac{S_\gamma(y^n)}{k^3}.
\end{equation}
The proof can be found in the appendix.

This way of encoding the class index is by no means the only
possibility but it will be seen to work sufficiently well, except for
one curious limitation: As a consequence of modeling both the
informative coefficients and the noise by densities from the same
Gaussian model, the code length formula approaches the same value as
$k$ approaches either zero or $n$, which actually are disallowed.
Hence, it may be that in cases where there is little information to
recover, the random fluctuations in the data may yield a minimizing
solution near $k=n$ instead of a correct solution near $k=0$. A
similar phenomenon has been demonstrated for ``saturated'' Bernoulli
models with one parameter for each observation~\cite{foster:2005}, and
resembles the inconsistency problem of BIC in Markov chain order
selection~\cite{csiszar:2000}: In all these cases pure random noise is
incorrectly identified as maximally regular data.  In order to prevent
this we simply restrict $k \leq .95 n$, which seems to avoid such
problems. A general explanation and solution for these phenomena would
be of interest\footnote{Perhaps a solution could be found in
algorithmic information theory (Kolmogorov complexity) and the concept
of Kolmogorov \emph{minimal} sufficient
statistic~\cite{vereshchagin:2004} which is the simplest one of many
equally efficient descriptions. However, for practical purposes, a
modification of the concept is needed in order to account for the
fluctuations near the extremes, which are succumbed by the constant
$\mathcal{O}(1)$ terms in algorithmic information theory.}.

\subsection{Subband Adaptation}
\label{sec:subband}

\begin{figure}
\hspace*{-0.04\columnwidth}\includegraphics[width=1.07\columnwidth]{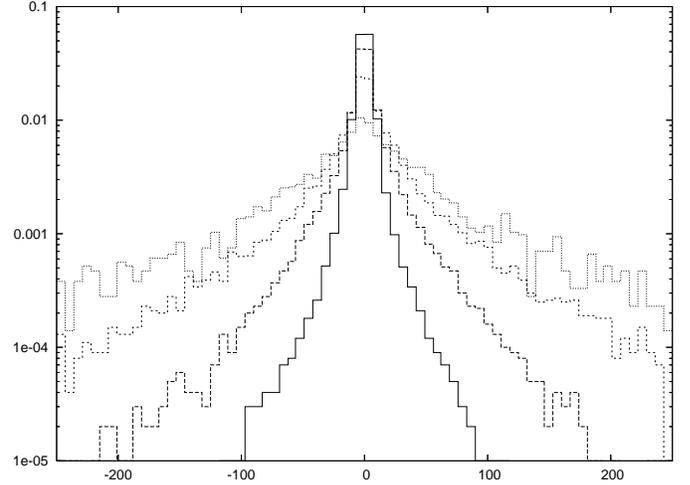}
\caption{Log-scale representation of the empirical histograms of the
wavelet coefficients on dyadic levels 6--9 for the Boat image (see
Sec.~\ref{sec:results} below).  Finer levels have narrower (more
sparse) distributions than coarser levels; the finest level (9) is
drawn with solid line.}
\label{fig:histo}
\vspace{3mm}
\end{figure}

It is an empirical fact that for most natural signals the
coefficients on different subbands corresponding to different
frequencies (and orientations in 2D data) have different
characteristics. Basically, the finer the level, the more sparse the
distribution of the coefficients, see Fig.~\ref{fig:histo}. (This is
not the case for pure Gaussian noise or, more interestingly, signals
with fractal structure~\cite{mallat:1989}.)  Within the levels the
histograms of the subbands for different orientations of 2D
transforms typically differ somewhat, but the differences between
orientations are not as significant as between levels.

In order to take the subband structure of wavelet transforms into
account, we let each subband $b \in \{1, \ldots, B\}$ have its own
variance, $\tau_b$. We choose the set of the retained coefficients
separately on each subband, and let $\gamma_b$ denote the set of the
retained coefficients on subband $b$, with $k_b := |\gamma_b|$. For
convenience, let $\gamma_0$ be the set of all the coefficients that
are not retained.  Note that this way we have $k_0 + \ldots + k_b =
n$. In order to encode the retained and the discarded coefficients
on each subband, we use a similar code as in the `flat' case
(Sec.~\ref{sec:encodemodel}). For each subband $1, \ldots, B$, the
number of nats needed is $\ln {n_b \choose k_b}$.

Ignoring again the constraint $\tau_b^2 + \sigma_N^2 \geq \sigma_N^2$,
the levels can be treated as separate sets of coefficients with their
own Gaussian densities just as in the previous subsection, where we had
two such sets.  The code length function, including the code length
for $\gamma$, becomes after Stirling's approximation to the Gamma
function and ignoring constants as follows:
\begin{equation}
\label{eq:levelwise}
  \sum_{b=0}^B \left(
  \frac{k_b}{2} \ln \frac{S_{\gamma_b}(y^n)}{k_b}
  + \frac{1}{2} \ln k_b\right)
  +\sum_{b=1}^B 
  \ln {n_b \choose k_b}.
\end{equation}
The proof is omitted since it is entirely analogous to the proof of
Eq.~\eqref{eq:onelevel} (see the appendix), the only difference being
that now we have $B+1$ Gaussian densities instead of only two.
Notwithstanding the added code-length for the retained indices, for
the case $B=1$ this coincides with the original setting, where the
subband structure is ignored, Eq.~\eqref{eq:onelevel}, since we then
have $k_0=n-k_1$. This code can be extended to allow $k_b=0$ for some
subbands simply by ignoring such subbands, which formally corresponds
to reducing $B$ in such cases\footnote{In fact, when reducing $B$ the
constants ignored also get reduced. This effect is very small compared
to terms in~\eqref{eq:levelwise}, and can be safely ignored since
codes with positive constants added to the code lengths are always
decodable.}.

Finding the index sets $\gamma_b$ that minimize the NML code length
simultaneously for all subbands $b$ is computationally demanding.
While on each subband the best choice always includes some $k_b$
largest coefficients, the optimal choice on subband $b$ depends on
the choices made on the $B-1$ other subbands. A reasonable
approximate solution to the search problem is obtained by iteration
through the subbands and, on each iteration, finding the locally
optimal coefficient set on each subband, given the current solution
on the other subbands.  Since the total code length achieved by the
current solution never increases, the algorithm eventually
converges, typically after not more than five iterations.
Algorithm~1 in Fig.~\ref{fig:al1} implements the above described
method. Following established
practice~\cite{donoho-johnstone:1995,chang:2000}, all coefficients
are retained on the smallest (coarsest) subbands\footnote{We retain
all subbands below level 4, i.e., all subbands with 16 or less
coefficients. This has little effect to the present method, but
since it is important for other methods to which we compare,
especially SureShrink, we adopted the practice in order to
facilitate comparison.}.

\begin{figure}[t!]
\vspace{2mm}\begin{tabular}{ll}
\multicolumn{2}{l}{\textsc{Algorithm 1.}}\\
\multicolumn{2}{l}{Input: signal $y^n$}\\\hline\hline
0. & \rule{0ex}{2.4ex}set $\coeff^n \leftarrow \calW^Ty^n$\\
1. & initialize $k_b = n_b$ for all $b \in \{1,\ldots,B\}$\\
2. & do until convergence\\
3. & \hspace{5mm}for each $b \in \{B_0+1, \ldots, B\}$\\
4. & \hspace{10mm}optimize $k_b$ wrt.\ criterion~\eqref{eq:levelwise}\\
5. & \hspace{5mm}end\\
6. & end\\
7. & for each $i \in \{1, \ldots, n\}$\\
8. & \hspace{5mm}if $i \notin \gamma$ then set $\coeff_n \leftarrow 0$\\
9. & end\\
10. & output $\calW \coeff^n$
\end{tabular}\vspace{2mm}
\caption{Outline of an algorithm for subband-adaptive MDL
denoising. The coarsest $B_0$ subbands are not processed in the loop
of Steps 3--5.  In Step~8, the final model $\gamma$ is defined by the
largest $k_b$ coefficients on each subband $b$. A soft thresholding
variation to Step 8 is described in Sec.~\ref{sec:mixture}.}
\label{fig:al1}
\end{figure}

\subsection{Soft Thresholding by Mixtures}
\label{sec:mixture}

The methods  described above can be used to determine the MDL model,
defined by a subset $\gamma$ of the wavelet coefficients, that gives
the shortest description to the observed data.  However, in many
cases there are several models that achieve nearly as good a
compression as the best one. Intuitively, it seems then too strict
to choose the single best model and discard all the others. A
modification of the procedure is to consider a \emph{mixture}, where
all models indexed by $\gamma$ are weighted by Eq.~\eqref{eq:prior}:
$$
   f_\mix(y^n) := \sum_\gamma f_\NML(y^n\;;\;\gamma) \,\pi(\gamma).
$$ Such a mixture model is universal (see 
e.g.~\cite{grunwald:2005,grunwald:2007})
in the sense that with increasing sample size the per sample average
of the code length $-n^{-1} \ln f_\mix(y^n)$ approaches that of the
best $\gamma$ for all $y^n$.  Consequently, predictions obtained by
conditioning on past observations converge to the optimal ones
achievable with the chosen model class.  A similar approach with
mixtures of trees has been applied in the context of
compression~\cite{willems:1995}.

For denoising purposes we need a slightly different setting since we
cannot let $n$ grow. Instead, given an observed signal $y^n$,
consider another image $z^n$ from the same source. Denoising is now
equivalent to predicting the mean value of $z^n$. Obtaining
predictions for $z^n$ given $y^n$ from the mixture is in principle
easy: one only needs to evaluate a conditional mixture
$$\begin{aligned}
   f_\mix(z^n \mid y^n) &= {f_\mix(y^n, z^n) \over f_\mix(y^n)}\\
   &= \sum_\gamma f_\NML(z^n \mid y^n \;;\; \gamma)
   \,\pi(\gamma \mid y^n).\notag
\end{aligned}$$
with new updated `posterior'
weights for the models, obtained by multiplying the NML density by the
prior weights and normalizing wrt.\ $\gamma$:
\begin{equation}\label{eq:nmlweights}
   \pi(\gamma \mid y^n) := \frac{f_\NML(y^n \;;\; \gamma) \pi(\gamma)}
   {\sum_{\gamma'} f_\NML(y^n \;;\; \gamma') \pi(\gamma')}.
\end{equation}  
Since in the denoising problem we only need the mean value instead
of a full predictive distribution for the coefficients, we can
obtain the predicted mean as a weighted average of the predicted
means corresponding to each $\gamma$ by replacing the density
$f_\NML(z^n \mid y^n \;;\; \gamma)$ by the coefficient value
$\coeff_i = \coeff_i(y^n)$ obtained from $y^n$ for $i \in \gamma$
and zero otherwise, which gives the denoised coefficients
\begin{equation}
\label{eq:barbeta}
   \sum_\gamma c_i \,\indic_{i \in \gamma} \,
   \pi(\gamma \mid y^n)
   = c_i \sum_{\gamma \ni i} \pi(\gamma \mid y^n),
\end{equation}
where the indicator function $\indic_{i \in \gamma}$ takes value one
if $i \in \gamma$ and zero otherwise.
Thus the mixture prediction of the coefficient value is
simply $\coeff_i$ times the sum of the weights of the models where $i \in
\gamma$
with the weights given by Eq.~\eqref{eq:nmlweights}.

The practical problem that arises in such a mixture model is that
summing over all the $2^n$ models is intractable. Since this sum
appears as the denominator of~\eqref{eq:nmlweights}, we cannot
evaluate the required weights. We now derive a tractable
approximation. To this end, let $\gamma_1 \ldots \gamma_n$ denote a
model determined by $i \in \gamma$ iff $\gamma_i = 1$, and let
$\gamma_1 \ldots 1_i \ldots \gamma_n$ denote a particular one
with $\gamma_i = 1$. Also, let $\hat\gamma = \hat\gamma_1 \ldots
\hat\gamma_n$ be the model with maximal NML posterior
weight~\eqref{eq:nmlweights}. The weight with which each individual
coefficient contributes to the mixture prediction can be obtained from
\begin{align}
\label{eq:ri}
   r_i &:= \frac{\sum_{\gamma \ni i} \pi(\gamma \mid y^n)}
   {\sum_{\gamma \not\ni i} \pi(\gamma \mid y^n)}
   = \frac{\sum_{\gamma \ni i} \pi(\gamma \mid y^n)}
   {1-\sum_{\gamma \ni i} \pi(\gamma \mid y^n)}\notag\\
   &\Longleftrightarrow
   \sum_{\gamma \ni i} \pi(\gamma \mid y^n) =
   \frac{r_i}{1+r_i}.
\end{align}
Note that the ratio $r_i$ is equal to
$$
r_i =
\frac{\sum_{\gamma} \pi(\gamma_1 \ldots 1_i \ldots \gamma_n \mid y^n)}
{\sum_{\gamma'}\pi(\gamma'_1 \ldots 0_i \ldots \gamma'_n \mid y^n)}.
$$
This can be approximated by
$$
\frac{\sum_{\gamma} \pi(\gamma_1 \ldots 1_i \ldots \gamma_n \mid y^n)}
{\sum_{\gamma'}\pi(\gamma'_1 \ldots 0_i \ldots \gamma'_n \mid y^n)}
\approx
\frac{\pi(\hat\gamma_1 \ldots 1_i \ldots \hat\gamma_n \mid y^n)}
{\pi(\hat\gamma_1 \ldots 0_i \ldots \hat\gamma_n \mid y^n)} := \tilde r_i,
$$ which means that the exponential sums in the numerator and the
denominator are replaced by their largest terms assuming that forcing
$\gamma_i$ to be one or zero has no effect on the other components of
$\hat\gamma$. The ratio of two weights can be evaluated without
knowing their common denominator, and hence this gives an efficient
recipe for approximating the weights needed in Eq.~\eqref{eq:barbeta}.

\begin{figure}[t!]
\includegraphics[width=\columnwidth]{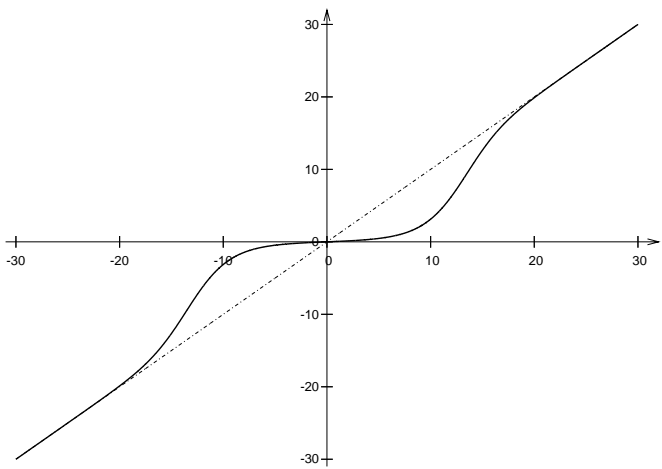}
\caption{The behavior of the soft thresholding method implemented by
Algorithm~2 for one of the subbands of the Boat image with no added
noise (see Sec.~\ref{sec:results}): the original wavelet coefficient
value $\coeff_i$ on the x-axis, and the thresholded value $\coeff_i
\,\tilde r_i/(1+\tilde r_i)$ on the y-axis.  For coefficients with
large absolute value, the curve approaches the diagonal (dotted
line). The general shape of the curve is always the same but the scale
depends on the data: the more noise, the wider the non-linear part.}
\label{fig:thres}
\end{figure}

Intuitively, if fixing $\gamma_i = 0$ decreases the posterior weight
significantly compared to $\gamma_i = 1$, the approximated value of
$r_i$ becomes large and the $i'$th coefficient is retained near its
maximum likelihood value $\coeff_i$. Conversely, coefficients that
increase the code length when included in the model are shrunk towards
zero. Thus, the mixing procedure implements a general form of `soft'
thresholding, of which a restricted piece-wise linear form has been
found in many cases superior to hard thresholding in earlier
work~\cite{donoho-johnstone:1994,chang:2000}. Such soft thresholding
rules have been justified in earlier works by their improved
theoretical and empirical properties, while here they arise naturally
from a universal mixture code. The whole procedure for mixing
different coefficient subsets can be implemented by replacing Step 8
of Algorithm~1 in Fig.~\ref{fig:al1} by the instruction
$$
  \mbox{set $\displaystyle c_i \leftarrow \coeff_i \,
  \frac{\tilde r_i}{1+\tilde r_i}$}
$$
where $\tilde r_i$ denotes the approximated value of $r_i$.  The
behavior of the resulting soft threshold is illustrated in
Fig.~\ref{fig:thres}.


\section{Experimental Results}
\label{sec:results}

\subsection{Data and Setting}

The effect of the three refinements of the MDL denoising method was
assessed separately and together on a set of artificial 1D
signals~\cite{donoho-johnstone:1995} and natural images\footnote{The
images were the same as used in many earlier papers, available at
\url{http://decsai.ugr.es/~javier/denoise/}.}  commonly used for
benchmarking. The signals were contaminated with Gaussian
pseudo-random noise of known variance $\sigma^2$, and the denoised
signal was compared with the original signal. The Daubechies D6 wavelet
basis was used in all experiments, both in the 1D and 2D
cases. The error was measured by the peak-signal-to-noise ratio
(PSNR), defined as
$$
   \text{\emph{PSNR}} := 
   10 \cdot \log_{10} 
   \left(\frac{\text{\emph{Range}}^2}{\text{\emph{MSE}}}\right),
$$ where $\text{\emph{Range}}$ is the difference between the maximum
and minimum values of the signal (for images $\text{\emph{Range}} =
255$); and $\text{\emph{MSE}}$ is the mean squared error.  The
experiment was repeated 15 times for each value of $\sigma^2$, and the
mean value and standard deviation was recorded.

The compared denoising methods were the original MDL
method~\cite{rissanen:2000} without modifications; MDL with the
modification of Sec.~\ref{sec:encodemodel}; MDL with the modifications
of Secs.~\ref{sec:encodemodel} and~\ref{sec:subband}; and MDL with the
modifications of Secs.~\ref{sec:encodemodel}, \ref{sec:subband} and
\ref{sec:mixture}. For comparison, we also give results for three
general denoising methods applicable to both 1D and 2D signals, namely
VisuShrink~\cite{donoho-johnstone:1994},
SureShrink~\cite{donoho-johnstone:1995}, and
BayesShrink~\cite{chang:2000}\footnote{All the compared
methods are available as a free package, downloadable at
\url{http://www.cs.helsinki.fi/teemu.roos/denoise/}. The package
includes the source code in C, using wavelet transforms from the Gnu
Scientific Library (GSL). All the experiments of
Sec.~\ref{sec:results} can be reproduced using the package.}.

\subsection{Results}
\label{subsec:results}

\begin{figure*}
\centering \includegraphics[width=2.3in]{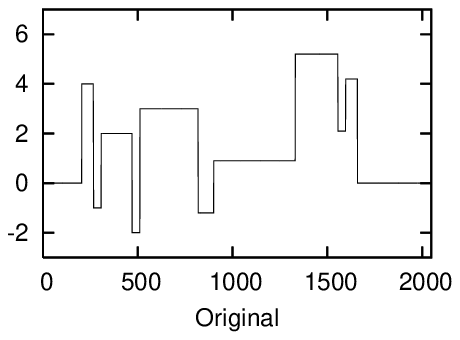}
\includegraphics[width=2.3in]{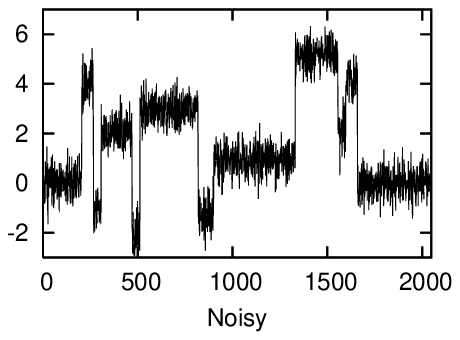}
\includegraphics[width=2.3in]{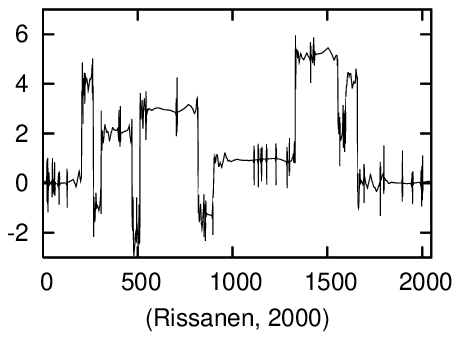}\\
\includegraphics[width=2.3in]{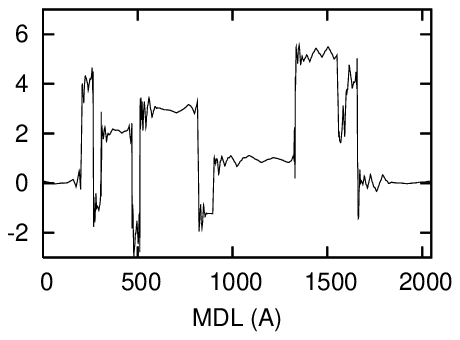}
\includegraphics[width=2.3in]{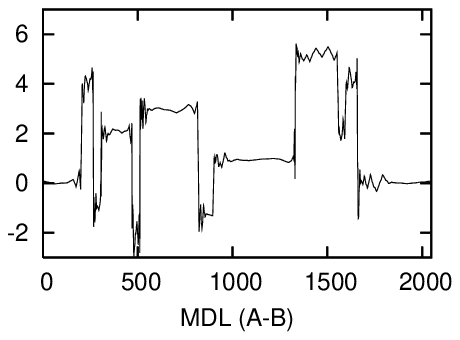}
\includegraphics[width=2.3in]{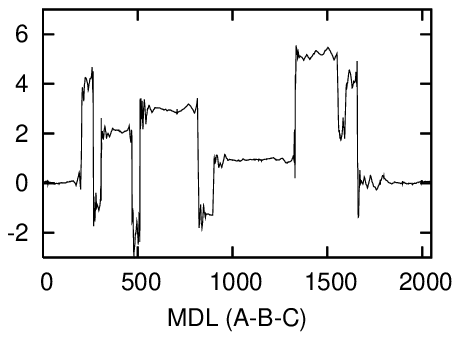}\\
\includegraphics[width=2.3in]{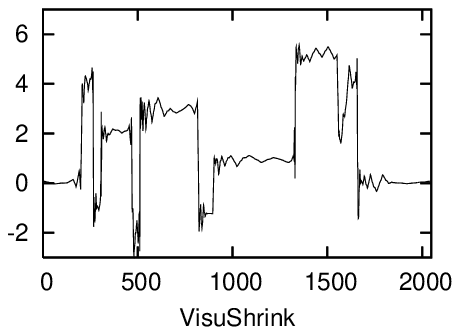}
\includegraphics[width=2.3in]{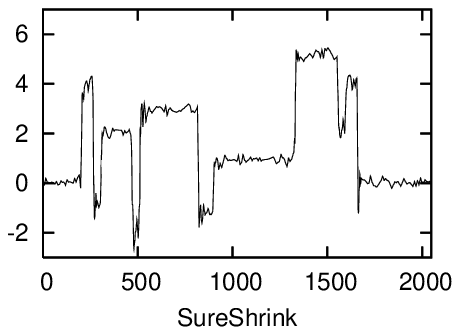}
\includegraphics[width=2.3in]{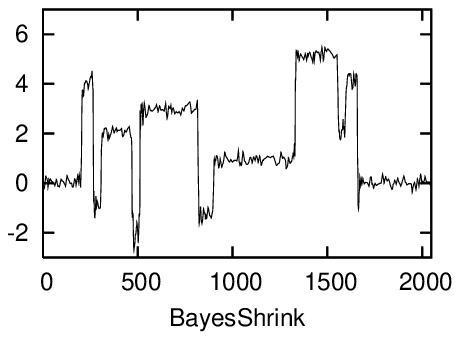}
\caption{Simulation Results. Panels from top to bottom, left to right:
Blocks signal~\cite{donoho-johnstone:1995}, sample size $n=2048$;
noisy signal, noise standard deviation $\sigma=0.5$, PSNR=23.2;
original MDL method~\cite{rissanen:2000}, PSNR=28.5; MDL with
modification of Sec.~\ref{sec:encodemodel}, PSNR=29.0; MDL with
modifications of Secs.~\ref{sec:encodemodel} and \ref{sec:subband},
PSNR=29.6; MDL with modifications of Secs.~\ref{sec:encodemodel},
\ref{sec:subband} and \ref{sec:mixture}, PSNR=30.1;
VisuShrink~\cite{donoho-johnstone:1994}, PSNR=28.6;
SureShrink~\cite{donoho-johnstone:1995}, PSNR=28.9;
BayesShrink~\cite{chang:2000}, PSNR=29.8. (Higher PSNR is better).}
\label{fig:blockdemo}
\end{figure*}

\begin{figure*}
\centering
\begin{tabular}{ccc}
\includegraphics[width=1.8in]{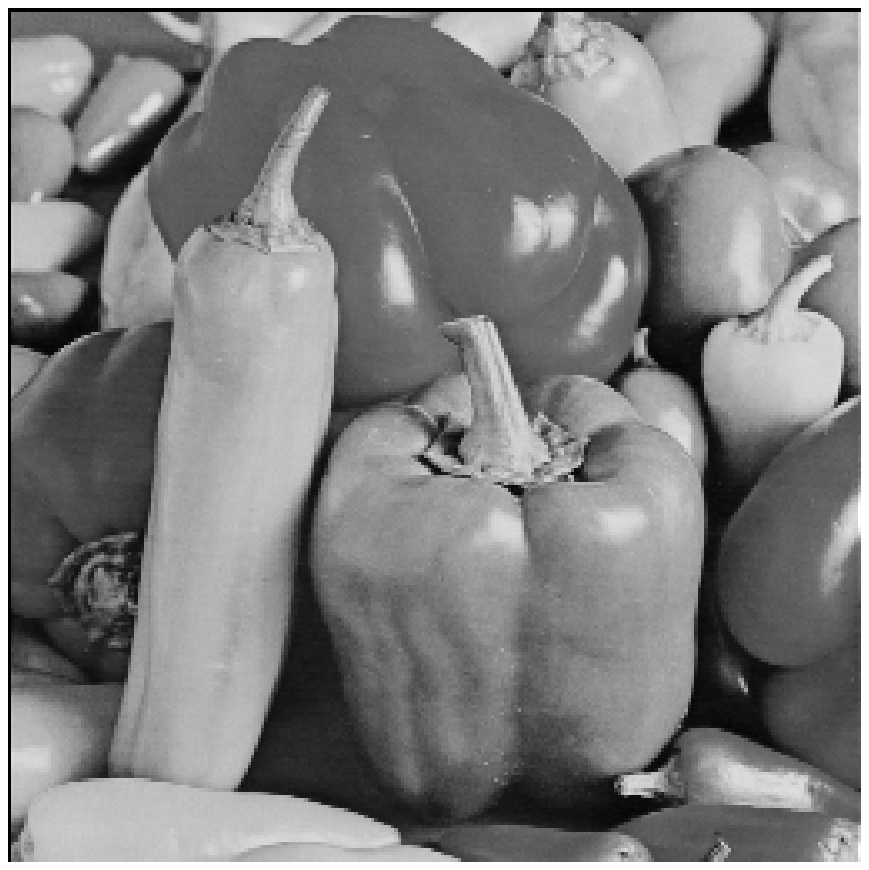} &
\includegraphics[width=1.8in]{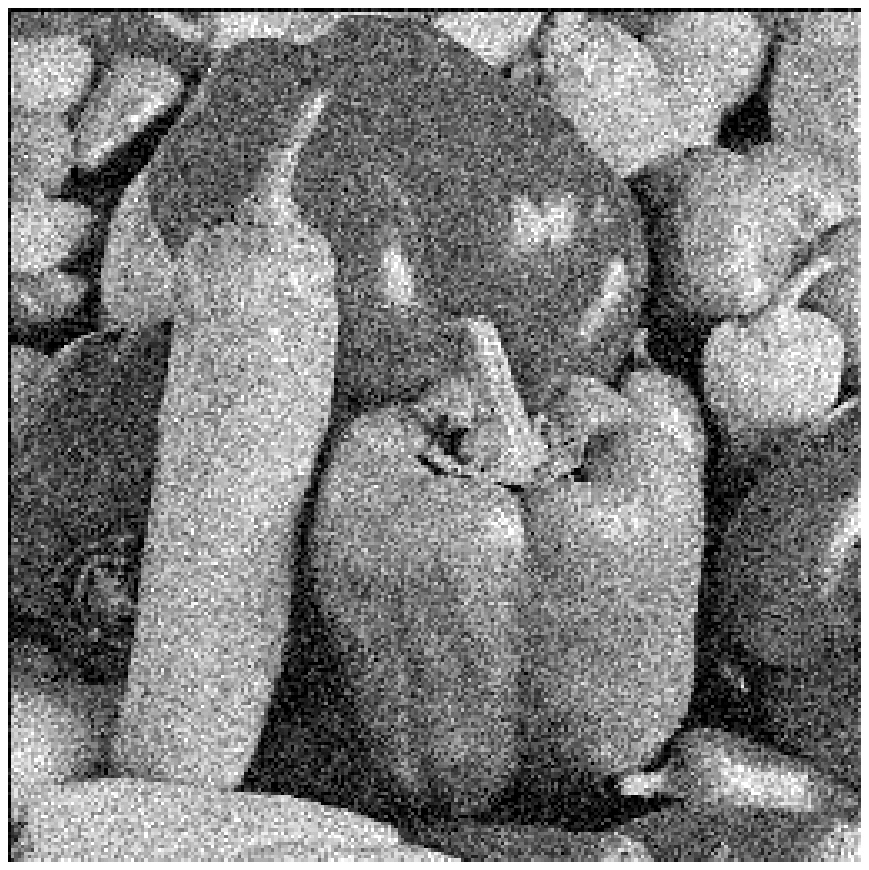} &
\includegraphics[width=1.8in]{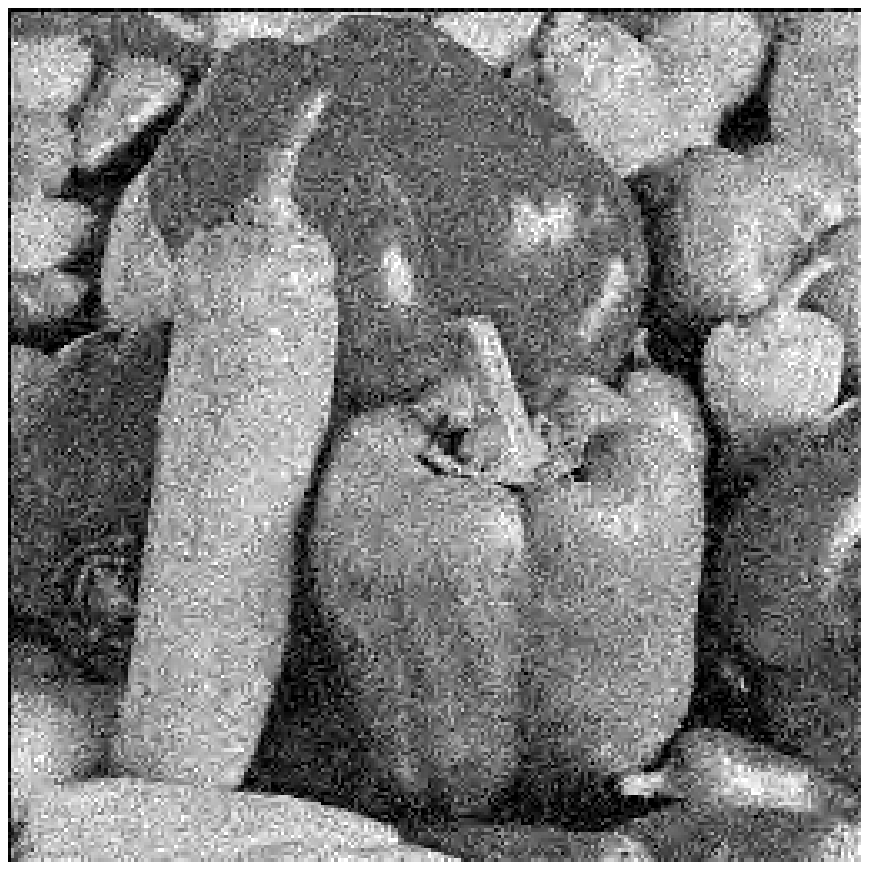} \\
\small Original & \small Noisy & \small (Rissanen, 2000) \vspace{1mm}\\
\includegraphics[width=1.8in]{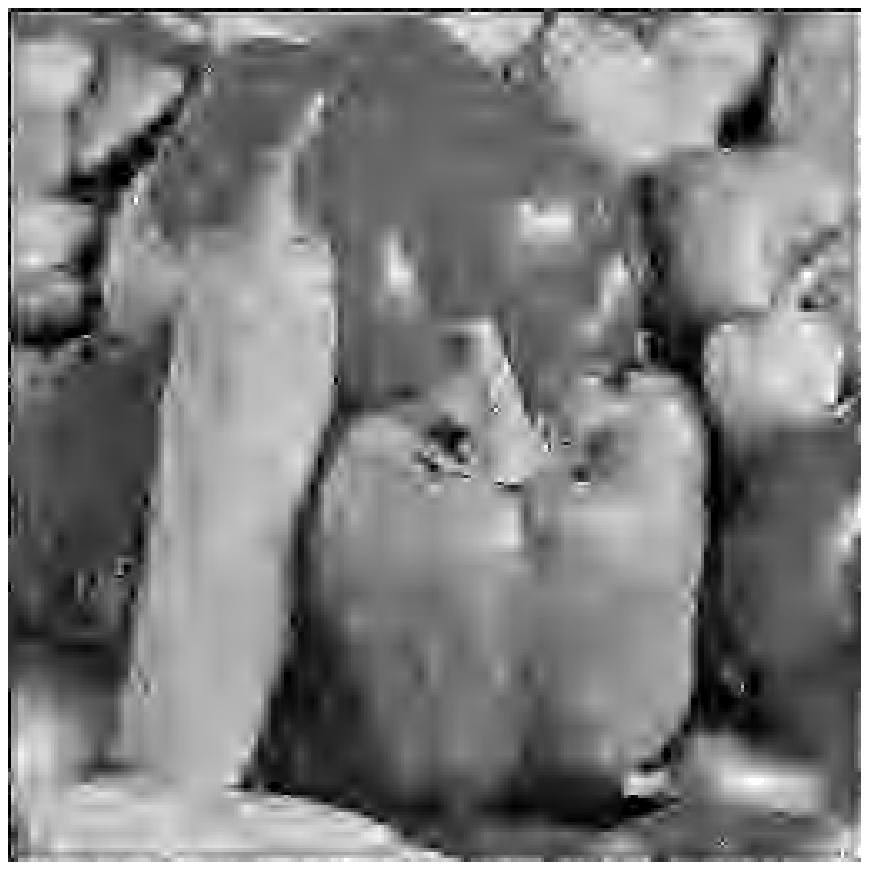} &
\includegraphics[width=1.8in]{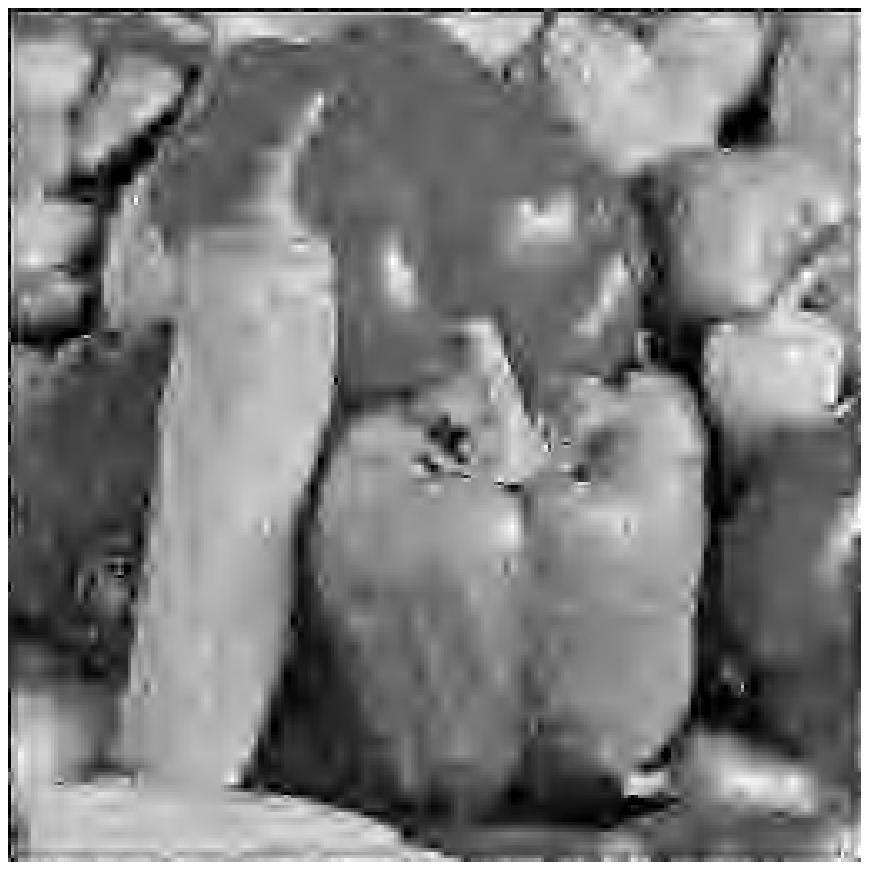} &
\includegraphics[width=1.8in]{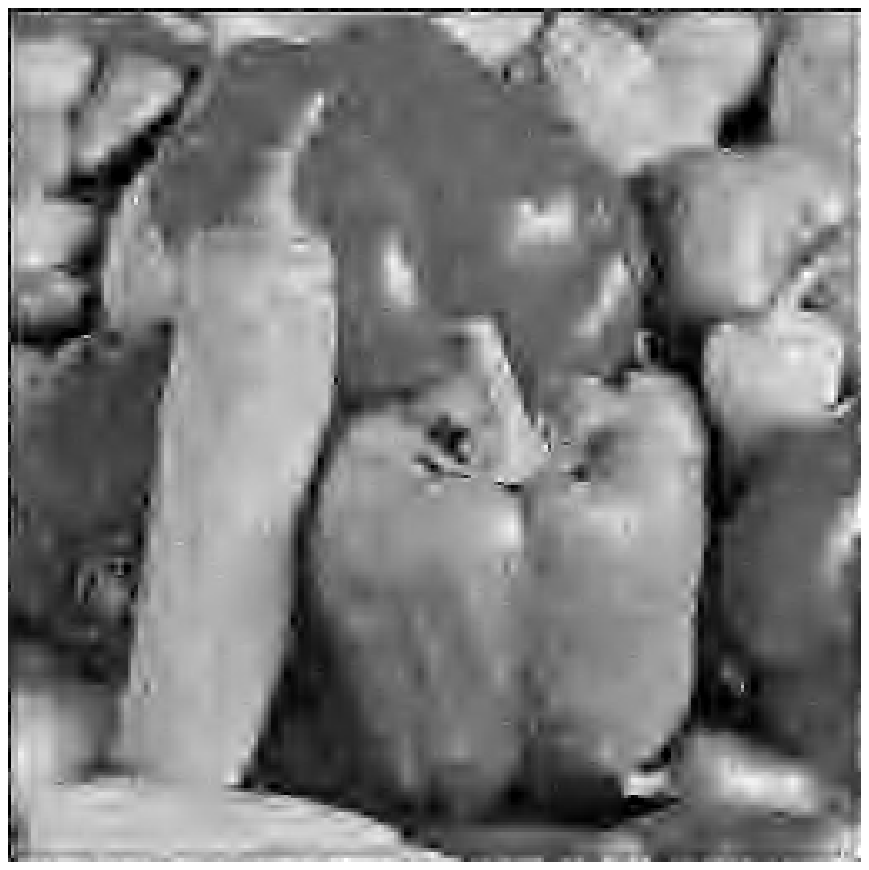} \\
\small MDL (A) & \small MDL (A-B) & \small MDL (A-B-C)\vspace{1mm} \\
\includegraphics[width=1.8in]{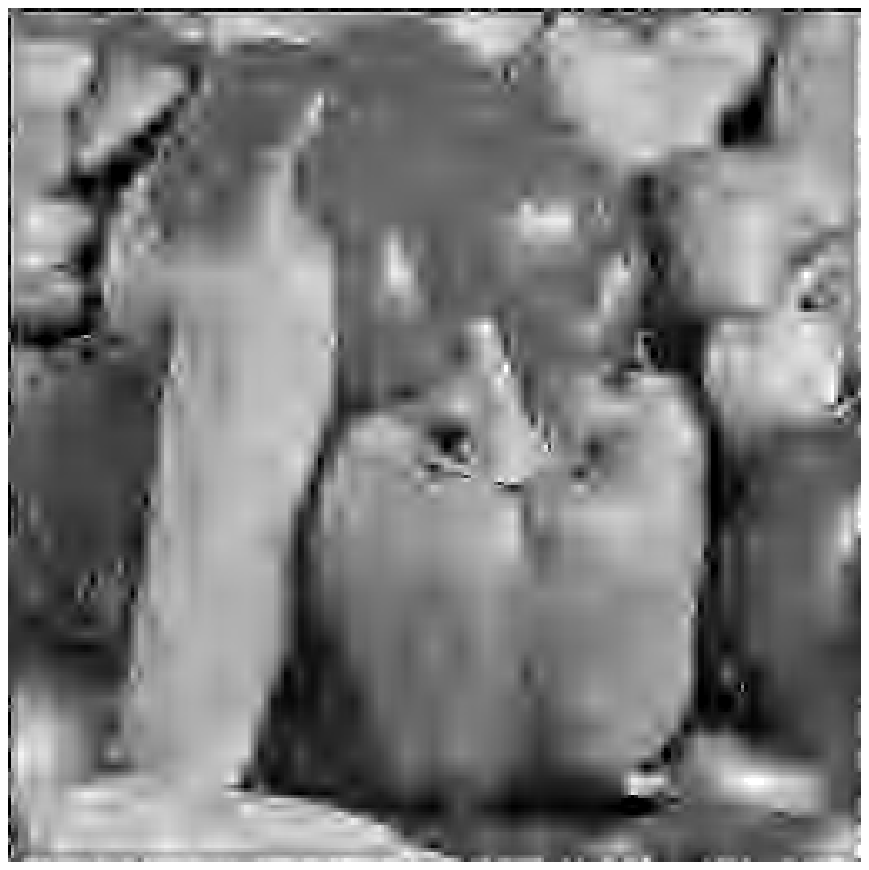} &
\includegraphics[width=1.8in]{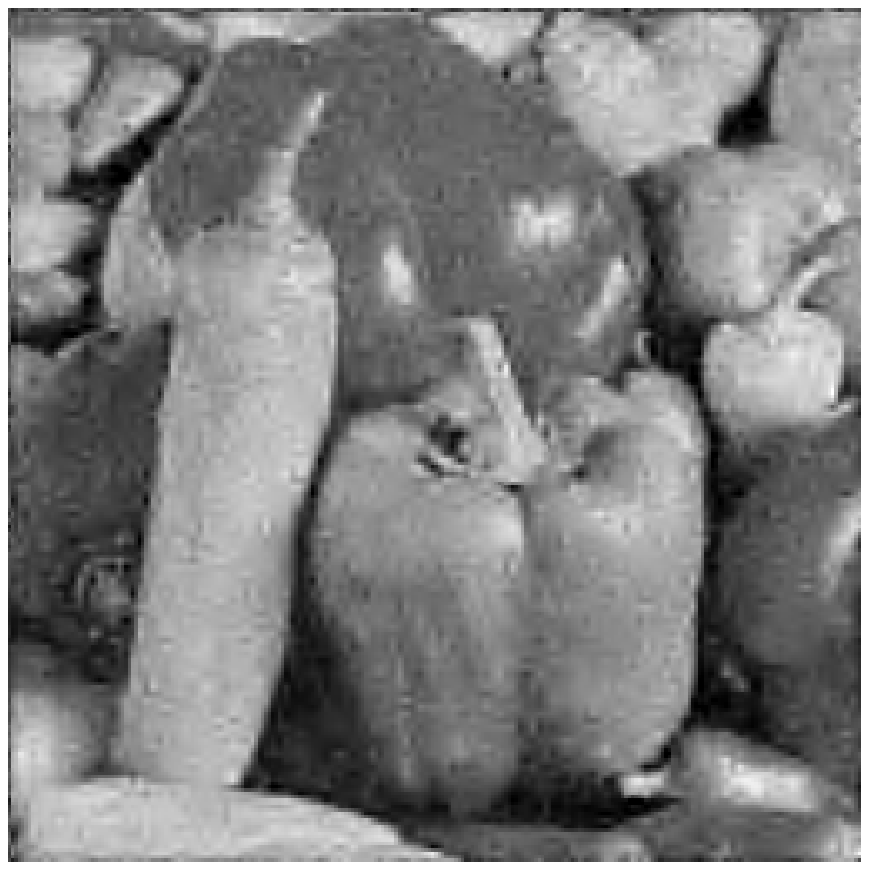} &
\includegraphics[width=1.8in]{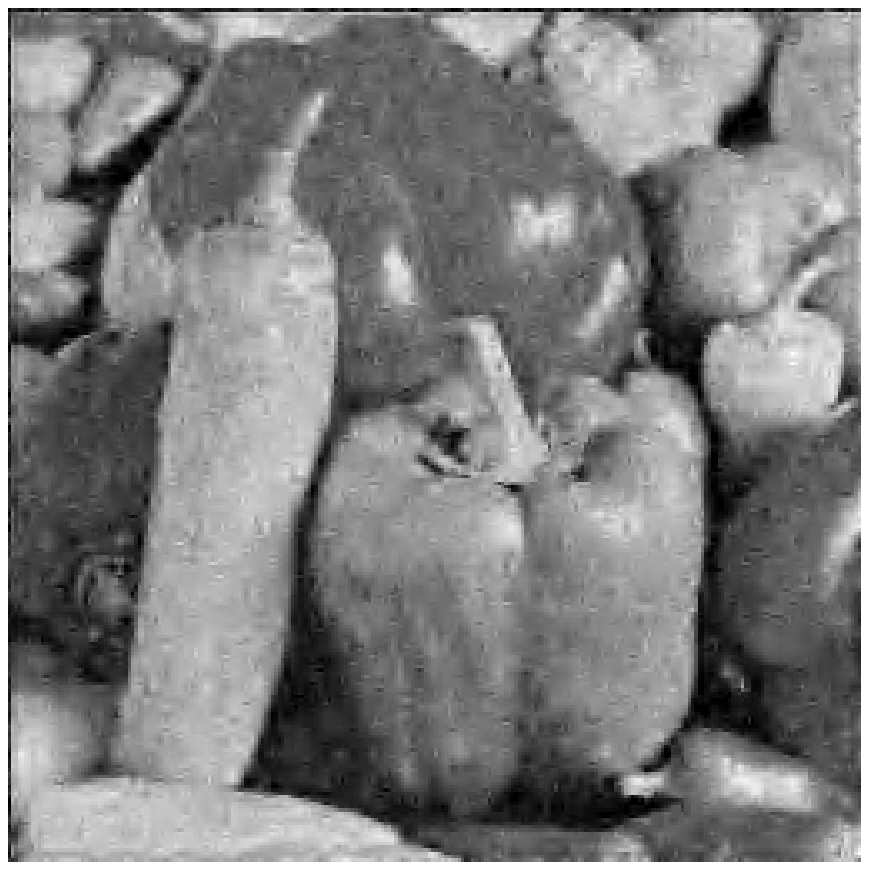} \\
\small VisuShrink & \small SureShrink & \small BayesShrink
\end{tabular}
\caption{Simulation Results. Panels from top to bottom, left to right:
Peppers image, $n = 256 \times 256$; noisy image, noise standard
deviation $\sigma=30$, PSNR=18.6; original MDL
method~\cite{rissanen:2000}, PSNR=19.9; MDL with modification of
Sec.~\ref{sec:encodemodel}, PSNR=23.9; MDL with modifications of
Secs.~\ref{sec:encodemodel} and \ref{sec:subband}, PSNR=24.9; MDL with
modifications of Secs.~\ref{sec:encodemodel}, \ref{sec:subband} and
\ref{sec:mixture}, PSNR=25.5; VisuShrink~\cite{donoho-johnstone:1994},
PSNR=23.2; SureShrink~\cite{donoho-johnstone:1995}, PSNR=24.6;
BayesShrink~\cite{chang:2000}, PSNR=25.9. (Higher PSNR is better).}
\label{fig:pepperdemo}
\end{figure*}

Figure~\ref{fig:blockdemo} illustrates the denoising results for the
\emph{Blocks} signal~\cite{donoho-johnstone:1995} with signal length
$n=2048$. The original signal, shown in the top-left display, is
piece-wise constant. The standard deviation of the noise is $\sigma =
0.5$.  The best method, having the highest $\text{\emph{PSNR}}$ (and
equivalently, the smallest $\text{\emph{MSE}}$) is the MDL method with
all the modifications proposed in the present work, labeled MDL
(A-B-C) in the figure. Another case, the \emph{Peppers} image with
noise standard deviation $\sigma = 30$, is shown in
Fig.~\ref{fig:pepperdemo}, where the best method is
BayesShrink. Visually, SureShrink and BayesShrink give a similar
result with some remainder noise left, while MDL (A-B-C) has removed
almost all noise but suffers from some blurring.

\begin{figure*}
\centering
{\small \sc Blocks (n=2048)}\\
\includegraphics[width=3.4in]{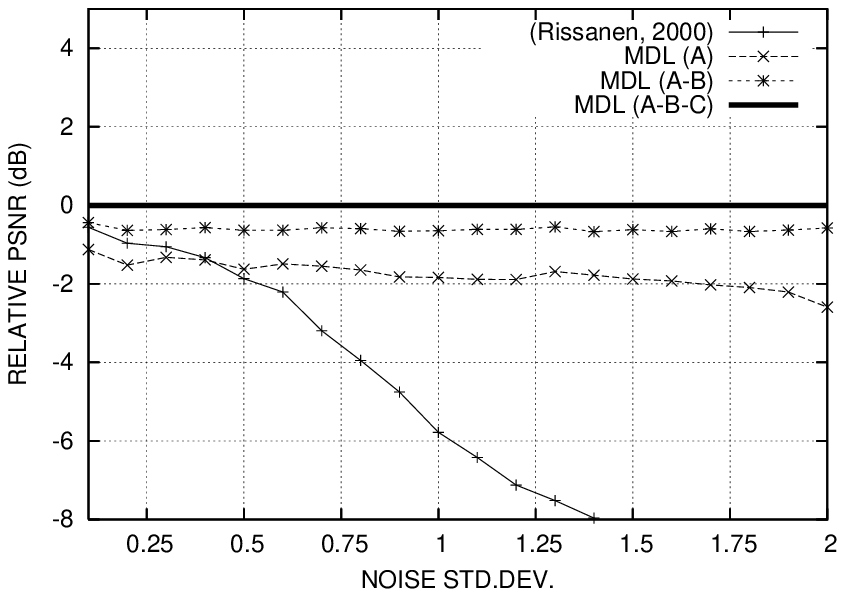}
\includegraphics[width=3.4in]{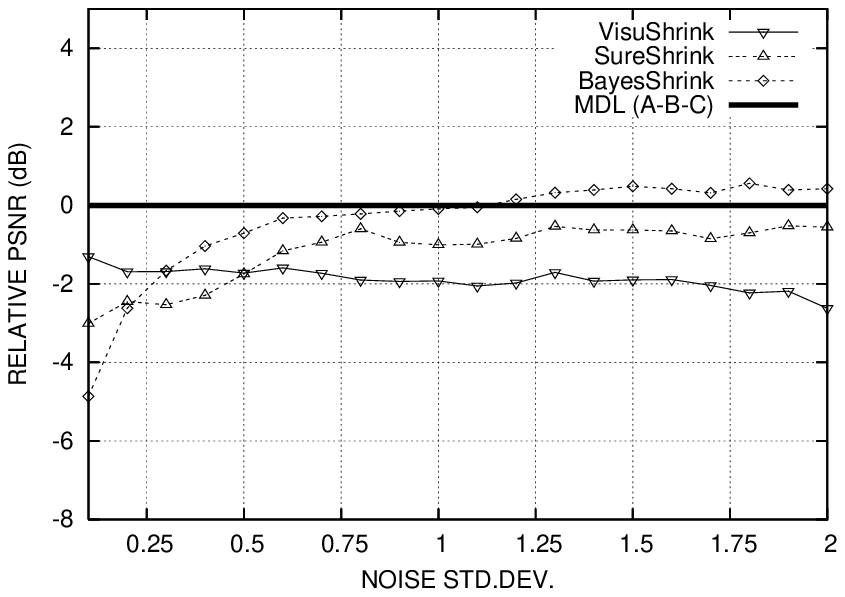}\\
\vspace{1mm}
{\small \sc Peppers}\\
\includegraphics[width=3.4in]{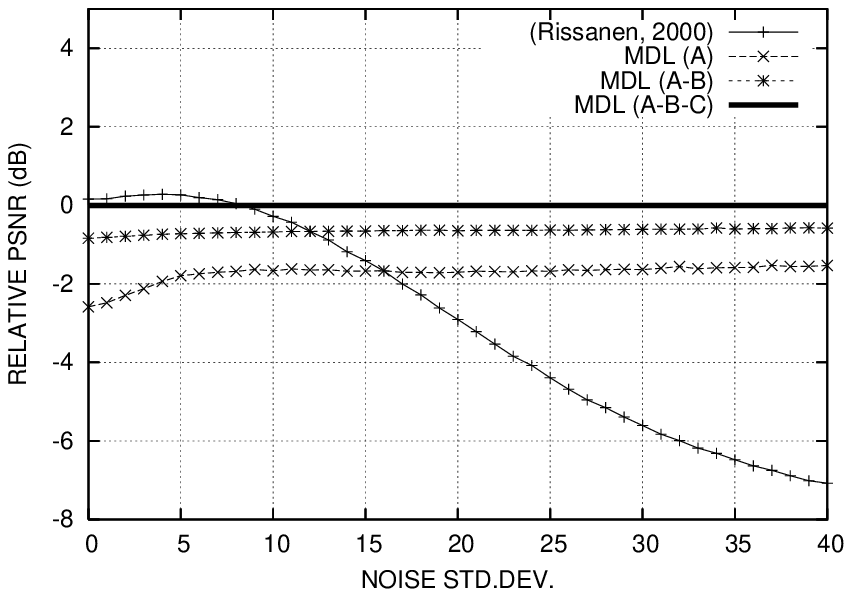}
\includegraphics[width=3.4in]{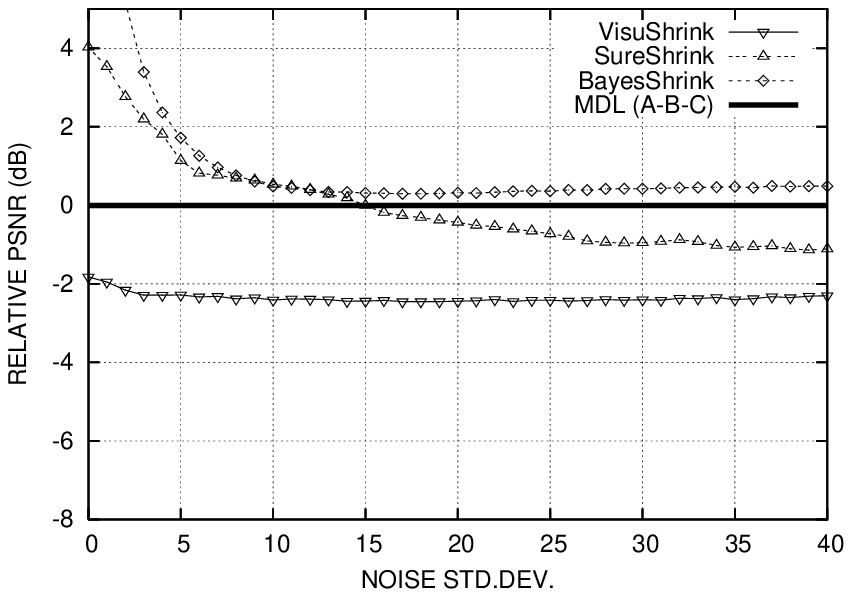}
\caption{Simulation Results. PSNR difference compared to the proposed
method (MDL with modifications of Secs.~\ref{sec:encodemodel},
\ref{sec:subband} and \ref{sec:mixture}), see
Figs.~\ref{fig:blockdemo} and~\ref{fig:pepperdemo}. Top row: Blocks
signal~\cite{donoho-johnstone:1995}, sample size $n=2048$. Bottom row:
Peppers image, $n=256 \times 256$. Left panels show the effect of each
of the three modifications in Sec.~\ref{sec:refined}; right panels
show comparison to VisuShrink~\cite{donoho-johnstone:1994},
SureShrink~\cite{donoho-johnstone:1995}, and
BayesShrink~\cite{chang:2000}.}
\label{fig:curves}
\end{figure*}

The relative performance of the methods depends strongly on the noise
level. Figure~\ref{fig:curves} illustrates this dependency in terms of
the relative PSNR compared to the MDL (A-B-C) method.  It can be seen
that the MDL (A-B-C) is uniformly the best among the four MDL methods
except for a range of small noise levels in the \emph{Peppers} case,
where the original method~\cite{rissanen:2000} is slightly better.
Moreover, it can be seen that the modifications of
Secs.~\ref{sec:subband} and~\ref{sec:mixture} improve the performance
on all noise levels for both signals. The right panels of
Fig.~\ref{fig:curves} show that the overall best method is
BayesShrink, except for small noise levels in \emph{Blocks}, where the
MDL (A-B-C) method is the best. This is explained by the fact that the
generalized Gaussian model used in BayesShrink is especially apt for
natural images but less so for 1D signals of the kind used in the
experiments.

The above observations generalize to other 1D signals and images as
well, as shown by Tables~\ref{tab:1d} and~\ref{tab:2d}. For some
1D signals (\emph{Heavisine}, \emph{Doppler}) the SureShrink method is
best for some noise levels. In images, BayesShrink is consistently
superior for low noise cases, although it can be debated whether the
test setting where the denoised image is compared to the original
image, which in itself already contains some noise, gives meaningful
results in the low noise regime. For moderate to high noise levels,
BayesShrink, MDL (A-B-C) and SureShrink typically give similar PSNR
output.

\begin{table*}
\renewcommand{\arraystretch}{1.25}
\caption{Numerical Results. The peak-signal-to-noise ratio for various 1D
signals, denoising methods, and noise levels. Columns: noise standard
deviation $\sigma$; PSNR for different methods (see
Figs.~\ref{fig:blockdemo} and~\ref{fig:pepperdemo}), best value(s) in
boldface; SD: standard deviation of all PSNR's for each value of
$\sigma$ over 15 repetitions.}
\label{tab:1d}
\centering
\begin{tabular}{rccccccccc}
\hline\hline
& (Rissanen, 2000) & MDL (A) & MDL (A-B) & MDL (A-B-C)
& VisuShrink & SureShrink & BayesShrink & & SD\\
\hline
 Blocks ($n=2048$)\\
$\sigma=0.1$ 
    & 44.4 & 43.8 & 44.5 & \bf 44.9 & 43.6 & 41.9 & 40.1 & $\pm$ & 0.32\\
0.5 & 28.9 & 29.1 & 30.1 & \bf 30.8 & 29.0 & 29.0 & 30.1 & $\pm$ & 0.42\\
1.0 & 20.4 & 24.4 & 25.5 & \bf 26.2 & 24.3 & 25.2 & 26.1 & $\pm$ & 0.44\\
1.5 & 15.0 & 21.6 & 22.8 & 23.4 & 21.5 & 22.8 & \bf 23.9 & $\pm$ & 0.34\\
2.0 & 11.7 & 19.6 & 21.6 & 22.2 & 19.5 & 21.6 & \bf 22.6 & $\pm$ & 0.45\\
\\
 Bumps ($n=2048$)\\
$\sigma=0.1$ 
    & 39.4 & 39.6 & 40.0 & \bf 40.7 & 39.2 & 38.8 & 38.3 & $\pm$ & 0.38\\
0.5 & 20.6 & 26.8 & 27.8 & \bf 28.4 & 26.1 & 27.2 & 28.0 & $\pm$ & 0.40\\
1.0 & 13.9 & 21.5 & 23.0 & 23.7 & 21.3 & 23.3 & \bf 24.0 & $\pm$ & 0.30\\
1.5 & 10.3 & 18.6 & 20.6 & 21.3 & 18.9 & 20.5 & \bf 21.9 & $\pm$ & 0.40\\
2.0 &  7.9 & 17.7 & 19.2 & 19.9 & 17.9 & 19.5 & \bf 20.3 & $\pm$ & 0.38\\
\\
 Heavisine ($n=2048$)\\
$\sigma=0.1$ 
    & 51.3 & 50.4 & 51.3 & \bf 51.9 & 51.1 & 48.8 & 48.1 & $\pm$ & 0.60\\
0.5 & 35.6 & 37.4 & 39.1 & \bf 39.5 & 37.7 & 38.3 & 38.9 & $\pm$ & 0.61\\
1.0 & 27.0 & 32.9 & 34.1 & 34.6 & 33.2 & \bf 34.7 & 34.1 & $\pm$ & 0.70\\
1.5 & 19.8 & 30.6 & 31.6 & 32.0 & 30.8 & \bf 32.3 & \bf 32.3 & $\pm$ & 0.91\\
2.0 & 15.4 & 28.1 & 30.5 & 31.0 & 28.2 & 31.2 & \bf 31.3 & $\pm$ & 1.02\\
\\
 Doppler ($n=2048$)\\
$\sigma=0.1$ 
    & 24.5 & 28.4 & 29.2 & \bf 29.8 & 28.3 & 28.6 & 29.5 & $\pm$ & 0.46\\
0.5 & 6.2 & 17.8 & 19.3 & 19.9 & 17.7 & 19.6 & \bf 20.3 & $\pm$ & 0.70\\
1.0 & 0.1 & 12.6 & 15.4 & 16.0 & 13.1 & 16.1 & \bf 16.2 & $\pm$ & 0.83\\
1.5 & -3.5 & 10.7 & 13.3 & 13.7 & 10.8 & \bf 14.0 & 13.9 & $\pm$ & 0.75\\
2.0 & -5.9 & 9.9 & 11.3 & 11.5 & 10.1 & \bf 12.2 & 11.8 & $\pm$ & 0.89\\
\hline
\end{tabular}
\end{table*}

\begin{table*}
\renewcommand{\arraystretch}{1.25}
\caption{Numerical Results. The peak-signal-to-noise ratio for various
images, denoising methods, and noise levels. Columns: noise standard
deviation $\sigma$; PSNR for different methods (see
Figs.~\ref{fig:blockdemo} and~\ref{fig:pepperdemo}), best value(s) in
boldface; SD: standard deviation of all PSNR's for each value of
$\sigma$ over 15 repetitions.}
\label{tab:2d}
\centering
\begin{tabular}{rccccccccc}
\hline\hline
& (Rissanen, 2000) & MDL (A) & MDL (A-B) & MDL (A-B-C) &
 VisuShrink & SureShrink & BayesShrink & & SD\\
\hline
 \multicolumn{1}{r}{Lena ($512 \times 512$)} \\
$\sigma=0$ 
   & 39.1 & 36.6 & 38.5 & 39.3 & 37.3 & 43.2 & \bf 46.9 & $\pm$ & --\\
10 & 31.6 & 30.8 & 31.8 & 32.4 & 30.1 & 32.8 & \bf 33.1 & $\pm$ & 0.02\\
20 & 25.0 & 27.8 & 28.8 & 29.4 & 27.1 & 29.5 & \bf 29.9 & $\pm$ & 0.03\\
30 & 19.8 & 26.0 & 27.1 & 27.6 & 25.4 & 27.8 & \bf 28.2 & $\pm$ & 0.03\\
40 & 16.7 & 24.9 & 26.0 & 26.5 & 24.3 & 26.4 & \bf 27.0 & $\pm$ & 0.04\\
\\
 \multicolumn{1}{r}{Boat ($512 \times 512$)} \\
$\sigma=0$ 
   & 36.2 & 33.2 & 35.1 & 35.9 & 32.9 & 39.2 & \bf 40.3 & $\pm$ & --\\
10 & 30.2 & 28.6 & 29.8 & 30.5 & 28.0 & 31.3 & \bf 31.7 & $\pm$ & 0.02\\
20 & 24.2 & 25.8 & 26.8 & 27.5 & 25.2 & 27.9 & \bf 28.3 & $\pm$ & 0.03\\
30 & 19.6 & 24.3 & 25.2 & 25.8 & 23.7 & 26.1 & \bf 26.5 & $\pm$ & 0.02\\
40 & 16.6 & 23.2 & 24.2 & 24.7 & 22.8 & 24.9 & \bf 25.3 & $\pm$ & 0.03\\
\\
 \multicolumn{1}{r}{House ($256 \times 256$)} \\
$\sigma=0$ 
   & 41.4 & 36.7 & 42.5 & 43.5 & 41.0 & 47.4 & \bf 54.2 & $\pm$ & --\\
10 & 31.4 & 30.7 & 31.5 & 32.1 & 30.2 & 32.5 & \bf 32.8 & $\pm$ & 0.06\\
20 & 24.7 & 27.3 & 28.1 & 28.7 & 26.8 & 28.7 & \bf 29.2 & $\pm$ & 0.05\\
30 & 19.7 & 25.4 & 26.4 & 27.0 & 24.9 & 26.9 & \bf 27.4 & $\pm$ & 0.06\\
40 & 16.7 & 24.2 & 25.2 & 25.7 & 23.7 & 25.4 & \bf 26.2 & $\pm$ & 0.07\\
\\
 \multicolumn{1}{r}{Peppers ($256 \times 256$)} \\
$\sigma=0$ 
   & 38.9 & 36.1 & 37.9 & 38.7 & 36.9 & 42.7 & \bf 51.2 & $\pm$ & --\\
10 & 30.7 & 29.3 & 30.3 & 31.0 & 28.6 & \bf 31.5 & \bf 31.5 & $\pm$ & 0.04\\
20 & 24.7 & 25.9 & 26.9 & 27.6 & 25.1 & 27.1 & \bf 27.9 & $\pm$ & 0.05\\
30 & 19.9 & 23.9 & 24.9 & 25.5 & 23.1 & 24.6 & \bf 25.9 & $\pm$ & 0.05\\
40 & 16.8 & 22.4 & 23.3 & 23.9 & 21.6 & 22.8 & \bf 24.4 & $\pm$ & 0.08\\
\hline
\end{tabular}
\end{table*}

\section{Conclusions}
\label{sec:conclusions}

We have revisited an earlier MDL method for wavelet-based denoising
for signals with additive Gaussian white noise. In doing so we gave an
alternative interpretation of Rissanen's renormalization technique for
avoiding the problem of unbounded parametric complexity in normalized
maximum likelihood (NML) codes. This new interpretation suggested
three refinements to the basic MDL method which were shown to
significantly improve empirical performance.

The most significant contributions are: i) an approach involving what
we called the \emph{extended model}, to the problem of unbounded
parametric complexity which may be useful not only in the Gaussian
model but, for instance, in the Poisson and geometric families of
distributions with suitable prior densities for the parameters; ii) a
demonstration of the importance of encoding the model index when the
number of potential models is large; iii) a combination of universal
models of the mixture and NML types, and a related predictive
technique which should also be useful in MDL denoising methods
(e.g.~\cite{saito:1994,antoniadis:1997,kumar:2006}) that are based on
finding a single best model, and other predictive tasks.

\appendices
\section{Postponed Proofs}

\emph{Proof of Eq.~\eqref{eq:onelevel}:} The proof of
Eq.~\eqref{eq:onelevel} is technically similar to the derivation of the
\emph{renormalized} NML model in~\cite{rissanen:2000}, which goes back
to~\cite{dom:1996}.  First note that due to
orthonormality, the density of $y^n$ under the extended model is
always equal to the density of $\coeff^n$ evaluated at $\calW^T
y^n$. Thus, for instance, the maximum likelihood parameters for data
$y^n$ are easily obtained by maximizing the density of $\coeff^n$ at
$\calW^T y^n$. The density of $\coeff^n$ is given by
\begin{equation}
\label{eq:lik}
f(\coeff^n \;;\; \sigma_I^2, \sigma_N^2) = \prod_{i \in
\gamma} \phi(\coeff_i \;;\; 0,\sigma_I^2) \prod_{i \notin
\gamma} \phi(\coeff_i \;;\; 0, \sigma_N^2),
\end{equation}
where $\phi(\cdot \;;\; \mu, \sigma^2)$ denotes a Gaussian density function
with mean $\mu$ and variance $\sigma^2$.

Let $S_\gamma(y^n)$ be the sum of squares of the
wavelet coefficients with $i \in \gamma$:
$$
   S_\gamma(y^n) := \sum_{i \in \gamma} \coeff_i^2.
$$ and let $S(y^n)$ denote the sum of all wavelet
coefficients.  With slight abuse of notation, we also denote these
two by $S_\gamma(\coeff^n)$ and $S(\coeff^n)$, respectively.
Let $k$ be the size of the set $\gamma$.

The likelihood is maximized by parameters given by
\begin{equation}
\label{eq:mlpar}
  \hat\sigma_I^2 = \frac{S_\gamma(y^n)}{k}, \quad
  \hat\sigma_N^2 = \frac{S(y^n) - S_\gamma(y^n)}{n-k}.
\end{equation}
With the maximum likelihood parameters (\ref{eq:mlpar}) the
likelihood (\ref{eq:lik}) becomes
\begin{equation}
\begin{aligned}
\label{eq:numerator}
(2\pi e)^{-n/2} \left(\frac{S_\gamma(y^n)}{k}\right)^{-k/2}
\left(\frac{S(y^n) - S_\gamma(y^n)}{n-k}\right)^{-\frac{n-k}{2}}.
\end{aligned}
\end{equation}
The normalization constant $C$ is also easier to evaluate by integrating
the likelihood in terms of $\coeff^n$:
\begin{equation}\begin{aligned}
\label{eq:rawnorm}
   C &= A \int \left(S_\gamma(\coeff^n)\right)^{-k/2}
\left(S(\coeff^n) - S_\gamma(\coeff^n)\right)^{-\frac{n-k}{2}} 
\,d\coeff^n,
\end{aligned}
\end{equation}
where $A$ is given by
$$
   A = (2\pi e)^{-n/2} k^{k/2} (n-k)^{\frac{n-k}{2}},
$$ 
and the range of integration $R$ is defined by requiring that the
maximum likelihood estimators~\eqref{eq:mlpar}
are both within the interval $[\sigma^2_{\min}, \sigma^2_{\max}]$. It
will be seen that the integral diverges without these bounds.  The
integral factors in two parts involving only the coefficients with $i
\in \gamma$ and $i \notin \gamma$ respectively.  Furthermore, the
resulting two integrals depend on the coefficients only through the
values $S_\gamma(\coeff^n)$ and $S(\coeff^n)-S_\gamma(\coeff^n)$, and
thus, they can be expressed in terms of these two quantities as the
integration variables -- we denote them respectively by $s_1$ and
$s_2$. The associated Riemannian volume elements are infinitesimally
thin spherical shells (surfaces of balls); the first one with
dimension $k$ and radius $s_1^{1/2}$, the second one with dimension
$n-k$ and radius $s_2^{1/2}$, given by
$$\begin{aligned}
   \frac{\pi^{k/2} s_1^{k/2-1}}
   {\Gamma(k/2)} \,ds_1, \quad
   \frac{\pi^{(n-k)/2} s_2^{(n-k)/2-1}}
   {\Gamma((n-k)/2)} \,ds_2.
\end{aligned}$$\\
Thus the integral in~\eqref{eq:rawnorm} is equivalent to
\begin{multline}
\int_{k\sigma^2_{\min}}^{k\sigma^2_{\max}}
   \frac{\pi^{k/2} s_1^{k/2-1}}{\Gamma(k/2)}
   s_1^{-k/2} \,ds_1 \notag\\
   \times
   \int_{(n-k)\sigma^2_{\min}}^{(n-k)\sigma^2_{\max}}
   \frac{\pi^{(n-k)/2} s_2^{(n-k)/2-1}}{\Gamma((n-k)/2)}
   s_2^{-(n-k)/2} \,ds_2\notag.
\end{multline}
Both integrands become simply of the form $1/x$ and hence, the value of
the integral is given by
\begin{equation}
\label{eq:integral}
   \frac{\pi^{n/2}}{\Gamma(k/2) \Gamma((n-k)/2)} \left(\ln
   \frac{\sigma^2_{\max}}{\sigma^2_{\min}}\right)^2,
\end{equation}
Plugging (\ref{eq:integral}) into (\ref{eq:rawnorm}) gives the value
of the normalization constant
$$
   C = \frac{k^{k/2} (n-k)^{(n-k)/2}}
   {(2e)^{n/2} \Gamma(k/2)\Gamma((n-k)/2)}
   \left(\ln \frac{\sigma^2_{\max}}{\sigma^2_{\min}}\right)^2.
$$
Normalizing the numerator (\ref{eq:numerator}) by $C$, and
canceling like terms finally gives the NML density:
\begin{multline}
   f_{\NML}(y^n) = \frac{\Gamma(k/2)\Gamma((n-k)/2)}
   {\pi^{n/2} (S_\gamma(y^n))^{k/2} (S(y^n)-S_\gamma(y^n))^{(n-k)/2}}\\
   \times
   \left(\ln \frac{\sigma^2_{\max}}{\sigma^2_{\min}}\right)^{-2},
\end{multline}
and the corresponding code length becomes
$$
\begin{aligned}
  -\ln f_{\NML}(y^n) &=
  \frac{k}{2} \ln S_\gamma(y^n)
  + \frac{n-k}{2} \ln (S(y^n)-S_\gamma(y^n))\\
  &- \ln \Gamma\left(\frac{k}{2}\right)
  - \ln \Gamma\left(\frac{n-k}{2}\right)\\
  &+ \frac{n}{2} \ln \pi
  + 2 \ln \ln \frac{\sigma^2_{\max}}{\sigma^2_{\min}}.
\end{aligned}
$$
Applying Stirling's
approximation
$$
   \ln \Gamma(z) \approx \left(z-\frac{1}{2}\right)\ln z - z +
   \frac{1}{2}\ln 2\pi,                                                       $$
to the Gamma functions yields now
$$\begin{aligned}
   -\ln f_\NML(y^n) &\approx
  \frac{k}{2} \ln S_\gamma(y^n)
  + \frac{n-k}{2} \ln (S(y^n)-S_\gamma(y^n))\\
  &- \left(\frac{k-1}{2}\right)\ln\left(\frac{k}{2}\right)
  + \frac{k}{2}\\
  &- \left(\frac{n-k-1}{2}\right)\ln\left(\frac{n-k}{2}\right)
  + \frac{n-k}{2}\\
  &- \ln 2\pi + \frac{n}{2} \ln \pi
  + 2 \ln \ln \frac{\sigma^2_{\max}}{\sigma^2_{\min}}.
\end{aligned}$$
Rearranging the terms gives the formula
\begin{multline}
   -\ln f_\NML(y^n) \approx
  \frac{k}{2} \ln \frac{S_\gamma(y^n)}{k}
  + \frac{n-k}{2} \ln \frac{S(y^n)-S_\gamma(y^n)}{n-k}\\
   +\frac{1}{2} \ln k(n-k) + \text{\emph{const}},
\end{multline}
where $\text{\emph{const}}$ is a constant wrt.\ $\gamma$, given by
$$
  \text{\emph{const}} = \frac{n}{2} \ln 2\pi e - \ln 4\pi +
  2 \ln \ln \frac{\sigma^2_{\max}}{\sigma^2_{\min}}.
$$

\qed
\pagebreak

\emph{Proof of Proposition~\ref{prop:ignore}:}
The maximum likelihood parameters~\eqref{eq:mlpar} may violate the
restriction $\sigma_I^2 \geq \sigma_N^2$ that arises from the
definition $\sigma_I^2 \eqdef \tau^2 + \sigma_N^2$.
The restriction affects range of integration in Eq.~\eqref{eq:integral}
giving the non-constant terms as follows
\begin{multline}
   \int_{k\sigma_{\min}^2}^{k\sigma_{\max}^2} 
   \left(\int_{(n-k)\sigma_{\min}^2}^{((n-k)/k)s_1} 
      s_1^{-1} s_2^{-1} \,ds_2\right) \,ds_1\\
   = \int_{k\sigma_{\min}^2}^{k\sigma_{\max}^2} s_1^{-1}
   (\ln s_1 - \ln k \sigma_{\min}^2) \,ds_1.
\end{multline}
Using the integral $\int s_1^{-1} \ln s_1 \,ds_1 = \frac{1}{2} (\ln
s_1)^2$ gives then
\begin{equation}
    \frac{1}{2}(\ln k\sigma_{\max}^2)^2
      - \frac{1}{2}(\ln k\sigma_{\min}^2)^2
   - \ln k\sigma_{\min}^2\left(\ln 
      \frac{\sigma_{\max}^2}{\sigma_{\min}^2}\right),
\label{eq:midterm}
\end{equation}
where the first two terms can be written as
$$ 
   \frac{1}{2} 
   \left(\ln k\sigma_{\max}^2 + \ln k\sigma_{\min}^2\right)
   \left(\ln \frac{\sigma_{\max}^2}{\sigma_{\min}^2}\right).
$$
Combining with the third term of~\eqref{eq:midterm} changes the plus into
a minus and gives finally
$$
      \frac{1}{2} 
   \left(\ln \frac{\sigma_{\max}^2}{\sigma_{\min}^2}\right)
   \left(\ln \frac{\sigma_{\max}^2}{\sigma_{\min}^2}\right),
$$ which is exactly half of the integral in Eq.~\eqref{eq:integral},
the constant terms being the same. Thus, the effect of the restriction
on the code length where the \emph{logarithm} of the integral is
taken, is one bit, i.e., $\ln 2$ nats.

\qed

\emph{Proof of Eq.~\ref{eq:withchoose}:} The relevant terms in the
code length $\ln {n \choose k}$, i.e.\ those depending on $k$, for the
index of the model class are
$$\begin{aligned}
   -\ln (k!(n-k)!) &= -\ln [k(k-1)!(n-k)(n-k)!] \\
   &= -\ln (k(n-k)) - \ln \Gamma(k) - \ln \Gamma(n-k),
\end{aligned}$$
which gives after Stirling's approximation (ignoring constant terms)
\begin{multline}
   -\ln (k(n-k)) - \left(k - \frac{1}{2}\right) \ln k + k\\
   - \left(n-k-\frac{1}{2}\right) \ln (n-k) + (n-k)\\
   = - \frac{k}{2} \ln k^2 - \frac{n-k}{2} \ln (n-k)^2 
   + \frac{1}{2}\ln k(n-k) + n .
\end{multline}
Adding this to Eq.~\ref{eq:onelevel} (without the constant $n$) gives
Eq.~\eqref{eq:withchoose}.

\qed




\section*{Acknowledgment}
The authors thank Peter Gr{\"u}nwald, Steven de Rooij, Jukka
Heikkonen, Vibhor Kumar, and Hannes Wettig for valuable comments.



%

\balance




\end{document}